\documentclass[10pt,a4paper]{article}

\def\heading #1{\bigbreak \begin{center} {\bf #1} \end{center}}

\title{Stability and production of positron-diatomic molecule complexes}

\author{Massimo Mella\ \\
Dipartimento di Chimica Fisica ed Elettrochimica,\\ Universita' degli Studi
di Milano, via Golgi 19, 20133 Milano, Italy\\
Electronic mail: Massimo.Mella@unimi.it\\
\and 
Dario Bressanini$^a$ and Gabriele Morosi$^b$\\
Dipartimento di Scienze Chimiche, Fisiche e Matematiche,\\
Universita' dell'Insubria, \\
via Lucini 3, 22100 Como, Italy\\
$^a$Electronic mail: Dario.Bressanini@uninsubria.it \\
$^b$Electronic mail: Gabriele.Morosi@uninsubria.it \\}

\begin{document}

\maketitle

\bigbreak
\begin{abstract}
The energies at geometries close to the equilibrium
for the e$^+$BeO and e$^+$LiF ground states were computed by means of diffusion 
Monte Carlo simulations. These results allow us to predict the 
equilibrium geometries and the vibrational frequencies for these exotic systems,
and to discuss their stability with respect to 
the various dissociation channels.
Since the adiabatic positron affinities were found to be smaller 
than the dissociation energies for 
both complexes, we propose these two molecules as possible candidates 
in the challenge to produce and detect stable positron-molecule systems.
\end{abstract}

\subsubsection*{PACS number(s): 36.10.-k, 02.70.Lq}

\pagebreak

Despite the wide diffusion of positron and positronium (Ps) based
analytical techniques to study solids ~\cite{weber}, 
polymers ~\cite{consolati},
solutions ~\cite{cast}, and organic molecules in the gas 
phase ~\cite{surko,schpsh}, a direct observation of the compounds between
the positron and an atom or a molecule is still lacking. 
In fact $\Gamma_{2\gamma}$
annihilation rate from positron annihilation life-time spectroscopy
and angular correlation annihilation radiation
are the only standard measurements carried out during the
interaction positron-matter.
The prediction of these observables is required to infer the formation
of the positronic compounds, a task that appears complex, especially for
heavy atoms and ions or large molecules, due to the high accuracy that is
needed for the wave function that describes the complexes.

The theoretical work on positron containing systems is scarce, 
and in our opinion this is due
to the difficulty in describing accurately the electron-positron correlation
using standard quantum chemistry methods like Self Consistent Field (SCF),
Configuration Interaction, and Coupled Cluster methods ~\cite{strasci}.

Two more approaches have been pursued during the last years, 
namely Density Functional Theory (DFT)~\cite{niem}
and variational calculations based on
Explicitly Correlated Gaussian (ECG) trial wave functions ~\cite{ryz,stras3}:
they also suffer from practical drawbacks.
Although DFT methods have a convenient scaling of the computational
cost versus the system complexity, the
exact exchange-correlation potential between electrons and
the correlation potential between electrons and positron are only approximately known.
As far as ECG wave functions are concerned, two groups 
~\cite{ryz,stras3} showed that
accurate results can be obtained even for positron containing systems.
Unfortunately, the ECG wave functions suffer from the fast increase of the
computational cost with the number of particles, therefore preventing their use
for medium and large systems. Nevertheless,
accurate results can be obtained employing the frozen-core
approximation for atoms and molecules ~\cite{ryz}.

In our ongoing project
to study positronic compounds as a way to understand matter-antimatter
interactions and to predict the existence of a bound state for positron-atom
or positron-molecule complexes [10-15],
~\nocite{maxpsh,maxanion,maxmole,maxold,maxcor,maxh2ps}
we employ the fixed node diffusion Monte Carlo (FN-DMC) method ~\cite{reybook}. 
This technique is known to be able to recover most of 
the correlation energy between electrons and between electrons and a positron
[10-15,17-19].
~\nocite{maxpsh,maxanion,maxmole,maxold,maxcor,maxh2ps,yoshi1,yoshi2,jiang}
Although FN-DMC is a powerful technique, it is not easy
to reduce the nodal error introduced by the fixed node approximation.
This result might be achieved in principle by employing more accurate 
trial wave functions or resorting to the nodal release technique, 
but both approaches do not easily apply to
large systems, i.e. more than ten electrons, due to their computational cost.
Nevertheless, the FN-DMC method has given accurate positron affinities, 
as well as electron affinities ~\cite{gabaf}, for systems up to 
twelve electrons, both atoms and molecules,
exploiting the cancellation of nodal errors ~\cite{maxmole}.

In the quest for stable positronic complexes, we studied the
potential surface for e$^+$LiH by FN-DMC calculations ~\cite{maxlihe+}
and found that the equilibrium distance and the 
vibrational transitions are different from those of LiH, opening the possibility
for a spectroscopic detection of this compound. However, 
the LiH adiabatic positron
affinity (APA) is larger than the dissociation energy (DE), and a third body
would be required to dissipate the excess energy. 
We suggested to start from a van der Waals complex of LiH
with a rare gas, and to attach the positron to this so that the rare gas
should dissipate the excess energy. Similar consideration can be extracted
from the work by Mitroy and Ryzhikh ~\cite{mitlih}, where they employed a full
non-adiabatic approach and ECG functions to establish the stability of e$^+$LiH.

To avoid this complex mechanism, in this Letter
we investigate other systems to see if we can find a molecule 
whose APA is smaller
than the DE, allowing the positron to be attached and to form
the complex without the intervention of a third body. If the spectroscopic
properties of this compound differ from those of the parent molecule, it 
could be a good candidate for experimental observation.

We have performed accurate calculations of the total energy 
for e$^+$BeO and e$^+$LiF systems at various internuclear distances
by means of FN-DMC. 
These results allow us to obtain the equilibrium distances for 
both molecules and to compute the vibrational frequencies.

In the FN-DMC algorithm we sample a distribution of configurations 
in 3N dimensional space that represents $\Psi_0 \Psi_T$, where $\Psi_0$ is 
the ground state wave function
having the same nodal surfaces of the trial wave function $\Psi_T$.
Using this distribution we obtain a MC estimate of the fixed node energy $E_0$
using the mixed estimator

\begin{equation}
\label{e+mole1}
E_0 = \frac{1}{N}\sum _{i=1}^{N} E_{eloc}({\mathbf R}_i) = 
\frac{1}{N} \sum_{i=1}^{N} \frac{H \Psi_T({\mathbf R}_i)}{\Psi_T({\mathbf R}_i)}
\end{equation}

\noindent
In our calculations the trial wave function $\Psi_T$ is

\begin{equation}
\label{e+mole2}
\Psi_{T} = Det\left| \phi_{\alpha} \right| Det\left| \phi_{\beta} \right|
e^{U\left( r_{\mu \nu} \right) } \Omega \left( {\mathbf r}_p, r_{p \nu} \right)
\end{equation}

\noindent
$\phi_{\alpha,\beta}$ are orbitals and $e^{U\left( r_{\mu \nu} \right) }$
is the electronic correlation
factor used by Schmidt and Moskowitz in their works on atoms 
and ions ~\cite{sch90,mosk1}.
In Eq. \ref{e+mole2}

\begin{equation}
\label{e+mole3}
\Omega\left( {\mathbf r}_p, r_{p \nu} \right) = \sum_{i=1}^{N_{terms}} c_{i}
\Phi_{i} \left( {\mathbf r}_p, r_{p \nu} \right)
\end{equation}
\noindent
where~\cite{morosi,mecor}

\begin{eqnarray}
\label{e+mole4}
\Phi_{i}\left( {\mathbf r}_p, r_{p \nu} \right) = f_i (\mathbf{r}_{p})
exp \left[ k_{i,1} \sum _{\nu=1} ^{N} r_{p \nu}
-\sum _{n=1} ^{N_{nuc}} k_{i,n+1}r_{p,n}\right]
\end{eqnarray}

\noindent
In this equation
$f_i (\mathbf{r}_a)$ is a function that contains explicitly the dependence
on the spatial coordinates of the positron and $\mathbf{k}_{i}$
is a vector of parameters for the i--th term of the linear expansion.

While the $\phi_{\alpha,\beta}$ orbitals were obtained by means of
standard SCF calculations on the parent neutral molecule,
the other parameters of $\Psi_T$ were optimized minimizing the 
variance of the local energy
using a fixed sample of configurations. Although this method produces 
wave functions whose properties are generally less accurate 
than those obtained by minimizing the energy ~\cite{roth},
it is much faster. 
Moreover, the FN-DMC energy value depends only on the location
of the nodal surfaces of the electronic part of the wave function,
so that it is not extremely important to have the best possible 
description of its positronic part. Nevertheless, if one is 
interested in properties different from
the energy, whose accuracy is strongly dependent on the quality of the trial
wave function (for example the $\delta(r_{+-}$)), a re-optimization of all the
wave function parameters is needed ~\cite{maxcor}.

All the FN-DMC simulations were carried out using a target population
of 5000 configurations and a time step of 0.001 hartree$^{-1}$. 
Few more simulations
employing a time step of 0.0005 hartree$^{-1}$ were run to check 
for the absence of the time step bias in the mean energy values.
The FN-DMC energy results for various internuclear distances of e$^+$LiF 
and e$^+$BeO are shown in Table \ref{tab1}.

We fitted these energy values by means of a second order polynomial
and computed equilibrium geometrical parameters and the 
fundamental vibrational wavenumber $\omega_e$ for the two complexes
e$^+$$^7$Li$^{19}$F and e$^+$$^9$Be$^{16}$O.
All the results are collected in Table \ref{tab2}.

Comparing the results in Table \ref{tab2} with the experimental values
~\cite{hub} for
$^7$Li$^{19}$F ($R_{e}$ = 2.955 bohr, $\omega_e$ = 910.34 cm$^{-1}$) and
$^9$Be$^{16}$O ($R_{e}$ = 2.515 bohr, $\omega_e$ = 1487.32 cm$^{-1}$), we note
that after the addition of the positron both molecules have 
larger equilibrium distances and vibrational frequencies. 
While the increase of $R_e$ is similar to the one we
found for e$^+$LiH ~\cite{maxlihe+} and can be rationalized 
invoking the repulsive interaction of the
positron with the nuclei, the increase of stiffness of the two bonds is an 
unexpected result. However, it must be pointed out that the computed frequencies
have an estimated statistical accuracy of the order of 10\%, and this
means that care must be taken in discussing the change of this 
property.

In a previous work ~\cite{maxmole} we computed the total energies
for LiF (-107.4069(9) hartree) and BeO (-89.7854(13) hartree)
at their equilibrium distances by means of FN-DMC.
Together with the $E_{min}$ values shown in Table \ref{tab2}, these 
energies allow us
to compute the adiabatic positron affinity (APA) for these two systems, 
namely 0.022(1) hartree for e$^+$LiF, and 0.025(2) hartree for  e$^+$BeO.
These two values are smaller than the APA for the e$^+$LiH (0.0366(1) hartree).
This result was already observed for the vertical PA ~\cite{maxmole}, 
and is in contrast with the fact that the dipole moment of LiF
($\mu =$6.33 Debye)  and BeO ($\mu =$6.26 Debye) are larger than 
the one of LiH ($\mu =$5.88 Debye) ~\cite{dip}.
This indicates that the dipole moment is not sufficient to predict
a qualitative trend in the PA, and that this value strongly depends
on the specific features of each molecule.

As far as the dissociation of these complexes is concerned, care must be 
taken in choosing balanced values for the energies of the
fragments for the possible dissociation channels.
For a positron-diatomic molecule complex e$^+$MX, 
where M=Li or Be and X=O or F, the possible fragmentations are 
e$^+$M + X, M$^+$ + PsX, M + e$^+$X, and PsM + X$^+$.
Although not all the energy values of the fragments are known, 
one can safely assume that the PsM + X$^+$ dissociation
pattern has the highest energy with respect to the other possibilities.
This is due firstly to the large ionization potential of X (0.5005 hartree
for O, and 0.6403 hartree for F) ~\cite{moore}, 
at least twice as large as the positronium
(Ps) ground state energy (-0.25 hartree);
secondly, to the usually small binding 
energy of Ps to metal atoms (for instance, the binding energy of Ps to
Li in the PsLi complex is just 0.01158 hartree ~\cite{ryz}).
Moreover, we believe it is reasonable to discard also the M + e$^+$X channel,
since the possibility of obtaining binding between e$^+$ and X is hindered
by the small polarizability of X. To support this conclusion,
we stress that even for HF and H$_2$O, that are both polar molecules, DMC
did not show binding with the positron ~\cite{maxmole}. 
Although this is not a proof,
it strongly suggests that e$^+$O and e$^+$F probably are not bound.
Accepting these conclusions, we are left only with e$^+$M + X and M$^+$ + PsX
as possible fragmentations. To compute the total energy for both channels
we use the ECG results for
e$^+$Li (-7.532323 hartree), e$^+$Be (-14.669042 hartree), Li$^+$ 
(-7.279913 hartree), and Be (-14.667355 hartree) ~\cite{ryz},
and the FN-DMC results for O (-75.0518(4) hartree), F (-99.7176(3) hartree)
~\cite{arne1}, 
PsO (-75.3177(5) hartree), and PsF (-100.0719(8) hartree) ~\cite{maxanion}. 
Moreover, we estimate the Be$^+$ energy (-14.3248 hartree)
subtracting the ionization potential (0.3426 hartree) 
~\cite{moore} to the total 
energy of Be. Using these results, we end up with an energy of -107.2499(3)
hartree for e$^+$Li and F, and an energy of -107.3518(8) hartree for Li$^+$ 
and PsF.
This fragmentation, similar to the one found for e$^+$LiH
(i.e. Li$^+$ and PsH), is primarily driven by the small value of the Li 
ionization potential. Differently, for e$^+$BeO we obtain -89.6642(5) hartree
for Be$^+$ and PsO, and -89.7208(4) hartree for e$^+$Be and O, so that the
most stable dissociation fragments present a positron-atom bound state.
Using the lowest energy dissociation threshold for the two systems one gets
a DE of
0.080(1) hartree for e$^+$LiF, and 0.090(2) hartree for e$^+$BeO.
Both these values are larger than the APA, and this fact means that 
the two positron-molecule complexes do not dissociate after positron
addition to the parent molecules.
This outcome is different from what we found for the addition
of e$^+$ to LiH, where the e$^+$LiH complex breaks up due to the excess
of the APA with respect to the DE ~\cite{maxlihe+}.
Therefore, it does not appear necessary for LiF and BeO to use a third body,
and a simple positron addition
will give birth to stable complexes in rotovibrational excited states.
As previously stated,
the possibility to produce these stable species could give the chance
to experimentally detect stable positron complexes. 
Roughly speaking, a mean lifetime on the order of $10^{-9}$ seconds is expected
for these systems, and this may be large enough to allow a
spectroscopical analysis in the reaction chamber by means of Fourier Transform
Infrared Spectroscopy if a sufficient concentration of e$^+$MX can
be produced, and if the frequency shift after positron attachment 
is large enough that the vibrational spectrum of the complex does
not overlap with the neutral molecule one.
Unfortunately the large uncertainty in $\omega_e$ does not allow
a quantitative prediction of this frequency shift.
Moreover, positrons having kinetic energy larger than the
difference DE-APA can open the various fragmentation channels
depending on the excess of their relative energies.  For instance, the collision
between positron and BeO can produce e$^+$Be and O as fragments,
so that the annihilation of e$^+$ with the electronic cloud of Be can be
directly recorded from the 2 $\gamma$ photons.
Moreover, it might be possible to detect the stable state of PsF, a system
that, differently from PsCl and PsBr, has not been prepared in solution
~\cite{cast}.

In conclusion, we have presented accurate APA and DE for e$^+$LiF and 
e$^+$BeO systems computed by means of FN-DMC. These results allow us to 
discuss possible mechanisms of formation for positron-molecule 
complexes by direct
attachment of e$^+$ to the molecules, and the possibility to produce e$^+$M
and PsX systems. It should be now interesting to compute the 
$\Gamma _{2\gamma}$ annihilation rate for these complexes, in order to
predict their mean lifetime after e$^+$ addition.
Unfortunately, more technical work on the method appears to be necessary
before these calculations can be carried out for these large systems.

\heading{ACKNOWLEDGMENTS}
Financial support by the Universita' degli Studi di Milano 
is gratefully acknowledged. 
The authors are indebted to the Centro CNR per lo Studio delle Relazioni
tra Struttura e Reattivita' Chimica for grants of computer time.

\clearpage

\begin{table}
\begin{center}

\begin{tabular}{llc}  \hline\hline
&R     &  $\langle E \rangle$ \\ \hline
e$^+$LiF&2.955 & -107.4243(8) \\
&3.200 & -107.4291(8) \\
&3.400 & -107.4249(10) \\
&3.500 & -107.4176(8) \\ \hline
e$^+$BeO& 2.30  & -89.7975(13) \\
&2.40  & -89.8089(15) \\
&2.51  & -89.8108(18) \\
&2.75  & -89.7998(14) \\
\hline \hline

\end{tabular}
\caption{Total energy at various internuclear distances. 
All values are in atomic units.}
\label{tab1}
\end{center}
\end{table}

\clearpage

\begin{table}
\begin{center}

\begin{tabular}{lcc}  \hline\hline
 & e$^+$$^7$Li$^{19}$F & e$^+$$^9$Be$^{16}$O \\ \hline
$E_{min}$ (hartree) & -107.429(1) & -89.8108(16) \\
$R_{e}$ (bohr)    & 3.18       & 2.53 \\
$\omega_e$ (cm$^{-1}$) & 1073        & 1537     \\
$R_{0}$ (bohr)& 3.20 &2.55   \\

\hline \hline

\end{tabular}
\caption{Equilibrium properties for e$^+$$^7$Li$^{19}$F and e$^+$$^9$Be$^{16}$O
}
\label{tab2}
\end{center}
\end{table}

\end{document}